\documentclass[pra,aps,twocolumn,showpacs, superscriptaddress]{revtex4-2}
\usepackage{amsmath,amssymb,graphicx}
\usepackage{float}
\usepackage{comment}
\usepackage{bm}
\usepackage{bbold}

\usepackage{graphicx}
\usepackage{dcolumn}
\usepackage{bm}
\usepackage{hyperref}
\usepackage{xcolor}
\usepackage{braket}
\usepackage{textcomp}
\usepackage{comment}
\usepackage{inputenc}

\begin{document}

\title{Non-linear-enabled localization in driven-dissipative photonic lattices}

\author{A. Muñoz de las Heras}
\email{alberto.munoz@iff.csic.es}
\affiliation{%
Institute of Fundamental Physics IFF-CSIC, Calle Serrano 113b, 28006 Madrid, Spain}%
\author{A. Amo}
\affiliation{%
Univ. Lille, CNRS, UMR 8523 – PhLAM – Physique des Lasers Atomes et Mol\'ecules, F-59000 Lille, France}%
\author{A. González-Tudela}
\affiliation{%
Institute of Fundamental Physics IFF-CSIC, Calle Serrano 113b, 28006 Madrid, Spain}%

\begin{abstract} 

Recent experimental work has demonstrated the ability to achieve reconfigurable photon localization in lossy photonic lattices by continuously driving them with lasers strategically positioned at specific locations. This localization results from the perfect, destructive interference of light emitted from different positions and, because of that, occurs only at very specific frequencies. Here, we examine this localization regime in the presence of standard optical Kerr non-linearities, such as those found in polaritonic lattices, and show that they stabilize driven-dissipative localization across frequency ranges significantly broader than those observed in the linear regime. Moreover, we demonstrate that, contrary to intuition, in most siutations this driven-dissipative localization does not enhance non-linear effects like optical bistabilities, due to a concurrent reduction in overall intensities. Nevertheless, we are able to identify certain parameter regions where non-linear enhancement is achieved, corresponding to situations where emission from different spots constructively interferes.

\end{abstract}

\date{\today}

\maketitle

\section{Introduction}
\label{sec:intro}

Localization of waves is a ubiquitous interference phenomenon present in various fields, including solid state physics~\cite{Anderson1958,Lee1985}, acoustics~\cite{Kirkpatrick1985,Hu2008}, and photonics~\cite{Schwartz2007,Levi2011,Goblot2020}. In the photonic context, beyond its fundamental interest, localization can be exploited to increase light-matter coupling strengths, with applications in quantum information~\cite{haroche13a}, quantum communication~\cite{kimble08a,Reiserer2015,Reiserer2022}, and quantum simulation~\cite{ritsch13a,CarusottoCiuti2013}. To achieve such photon localization, several strategies have been explored, including the use of highly reflective mirrors in vacuum~\cite{haroche13a}, exploiting the confinement of light in materials with high refractive indices~\cite{joannopoulos97a,ebbesen08}, employing photonic crystals with a non-trivial topology leading to localized modes at their boundaries~\cite{Ozawa2019,St-Jean2017}, and, perhaps the most exotic, engineering bound-states-in-the-continuum. The latter appear when the localized waves coexist with a continuum spectrum of propagating modes protected by symmetry or separability~\cite{vonNeumann1993,Hsu2016}, and they have recently been predicted to arise also in the many-body regime~\cite{sugimoto2023manybody,huang2023interactioninduced}. In all these cases, however, the spatial nature of the localization is fixed by design and cannot be easily altered once the structure is built, limiting its versatility.

A recent experimental work~\cite{Jamadi2022} has demonstrated a new way forward to achieve re-configurable photon localization in lossy photonic lattices by exploiting the interplay of continuous, local drivings and non-trivial energy dispersions. The idea consists in placing several coherent pumps in judiciously chosen positions so that the light emitted from them destructively interferes and localizes within the region between them, being able to obtain highly non-trivial spatial patterns through appropriate laser modulation~\cite{Gonzalez-Tudela_2022}. As so far considered, the method presents two limitations: first, it only works for specific laser frequencies where the perfect destructive interference takes place; second, it was designed and tested in the linear regime~\cite{Jamadi2022,Gonzalez-Tudela_2022}, so that whether it works in the presence of non-linearities is still an open question.

In this work, we address both limitations by investigating driven-dissipative localization in photonic lattices with Kerr non-linearities~\cite{Butcher2008,CarusottoCiuti2013}. Our analysis reveals two counter-intuitive conclusions: First, non-linearities do not hinder driven-dissipative localization despite the non-homogeneous spatial distribution of the localization; in fact, they enable its existence across a broader range of frequencies than in purely linear lattices. Second, in spite of such localization, non-linear effects, like optical bistabilities, are weaker than in non-localized situations due to the smaller overall intensity in the former case. On the contrary, our study suggests that it is the opposite driving regime, in which the emission between the coherent pumps interferes constructively, the one that leads to an enhancement of non-linear optical effects.

The manuscript is organized as follows. In Sec.~\ref{sec:model}, we introduce the driven-dissipative setups considered throughout this manuscript, along with the theoretical tools used to model them. Secs.~\ref{sec:linear}-\ref{sec:1d non-linear} focus on a one-dimensional lattice with first-neighbor couplings. In Sec.~\ref{sec:linear}, we characterize the phenomenon of driven-dissipative localization in the linear regime, providing analytical expressions for the local and total intensities, which, to our knowledge, have not been presented elsewhere. Then, in Sec.~\ref{sec:1d non-linear}, we analyze the non-linear regime, concentrating on the stability of the localization and the potential enhancement or diminishment of optical bistability. Finally, in Sec.~\ref{sec:2d}, we demonstrate the generality of our results by considering a two-dimensional square lattice, and we summarize our conclusions and outlook in Sec.~\ref{sec:discussion}.

\section{Driven-dissipative photonic lattices}
\label{sec:model}

\begin{figure}[tb]
    \centering
    \includegraphics[width=\linewidth]{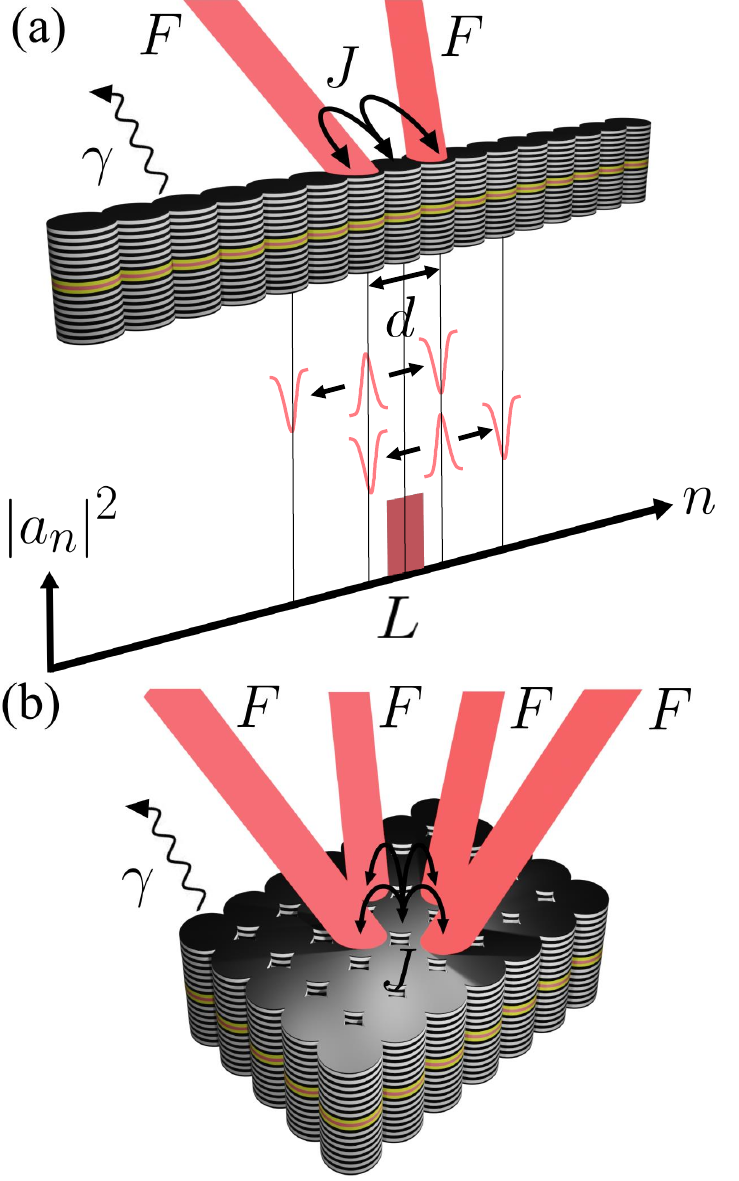}
    \caption{(a) Scheme of a one-dimensional lattice featuring first-neighbor couplings $J$, photon cavity decay rate $\gamma$, and driven by two coherent pumps with the same amplitude $F$. Localization takes place between two sites separated a distance $d$ when there is a destructive interference between the coherent light injected at the pumped sites. For instance, when $d=2$ this results in a single site labelled $n=L$ concentrating all the intensity in the lattice $|a_L|^2$. (b) An analog scheme for a two-dimensional square lattice. In this case, lasers of equal amplitude $F$ pump four sites of the lattice encircling a single site in which localization takes place.}
    \label{fig:scheme}
\end{figure}

\begin{figure*}[tb]
    \centering
    \includegraphics[width=\linewidth]{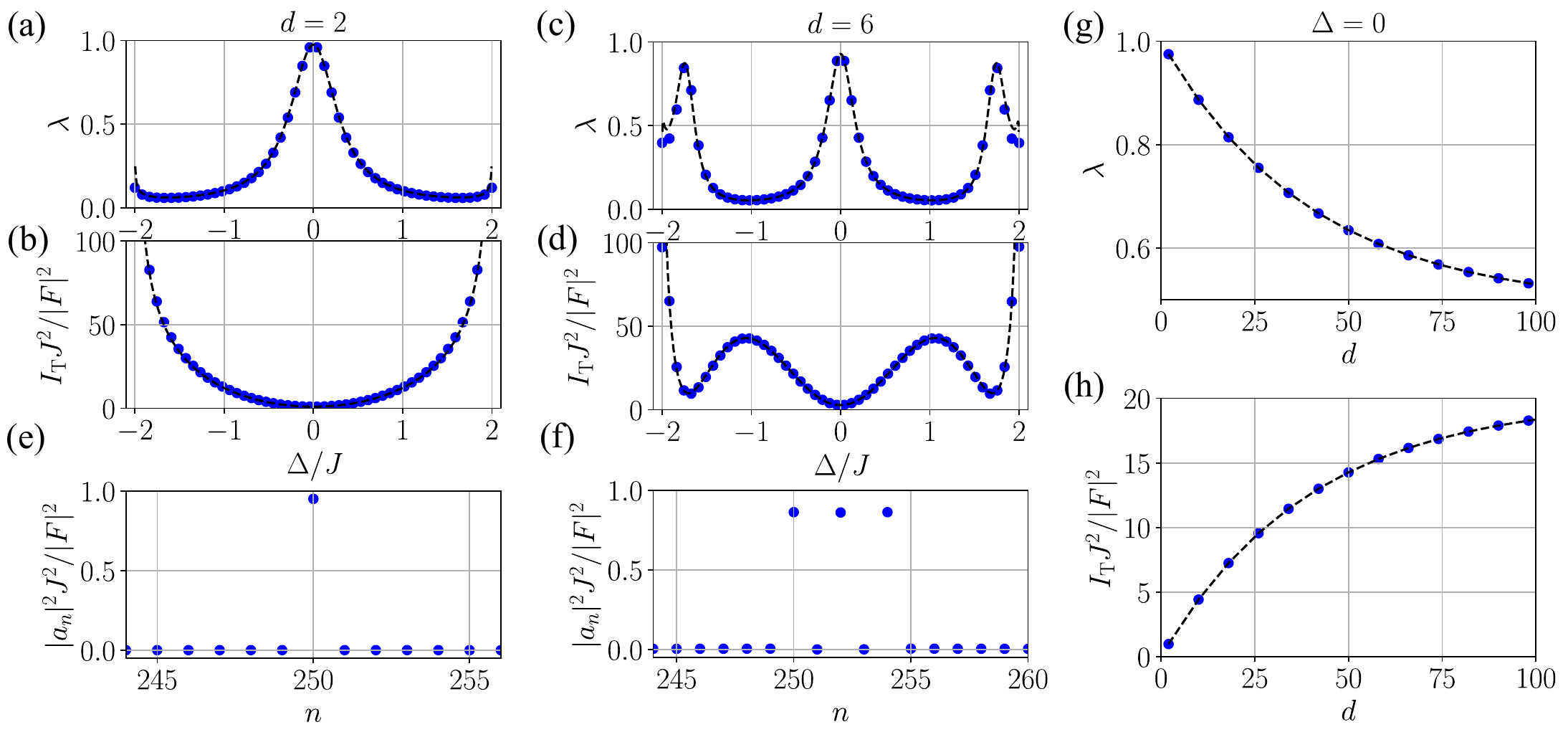}
    \caption{Analysis of localization in a linear ($U=0$) one-dimensional photonic lattice with first-neighbor couplings featuring two local pumps with equal amplitude. (a,b) [(c,d)] Localization $\lambda$ and total intensity in the lattice $I_{\rm T}$ (in units of the tunneling $J$ and the pump intensity $|F|^2$) as a function of the detuning $\Delta$ for an inter-pump distance $d=2$ [$d=6$]. (e) [(f)] Intensity $|a_n|^2$ in each site $n$ of the lattice for $\Delta=0$ and $d=2$ [$d=6$]. (g,h) $\lambda$ and $I_{\rm T}$ for a fixed $\Delta=0$ and as a function of $d$. In all panels, blue circles are the results of numerical simulations to find the steady state of Eq.~\eqref{eq:coupled mode}, while dashed lines correspond to the analytical results obtained with Eqs.~(\ref{eq:analytic intensity}-\ref{eq:analytic total intensity}). Parameters: cavity decay rate $\gamma = 5 \times 10^{-2} J$, $N=500$ sites with open boundary conditions, $n_1=250$.
    }
    \label{fig:linear}
\end{figure*}

In this work, we study the driven-dissipative steady-states of one (1D) and two-dimensional (2D) photonic lattices like the ones depicted in Fig.~\ref{fig:scheme}(a) and (b), respectively. The temporal dynamics and stationary states of these systems are described by the following set of coupled-mode equations governing the amplitude of the electric field at each site $a_n$~\cite{Fan2003,walls2008quantum}:
\begin{align}
    i\dot{a}_n &= \omega_a a_n
    - i\gamma a_n
    + U|a_n|^2 a_n
    - \sum_{\langle m\neq n \rangle} J_{n,m} a_{m}
    \nonumber\\&
    + F_n e^{-i\omega_{\rm P}t}
    ,
    \label{eq:coupled mode}
\end{align}
where the latin letters $n,m$ indicate the lattice site indices running from $1,\dots,N$ and from $1,\dots,N^2$ in the 1D and 2D cases, respectively, where $N$ is the system size along one dimension (throughout the paper we consider lattices with open boundary conditions); each site of the lattice corresponds to a single-mode resonator of bare resonance frequency $\omega_a$ and decay rate $\gamma$, which we assume to be equal for every site; $J_{n,m}$ is the tunneling rate between sites $m$ and $n$ (taken as a real number, since in this work we are not dealing with gauge fields~\cite{Ozawa2019}), which we restrict here to first-neighbors terms (in which $J_{\langle n,m \rangle}\equiv J$); $F_n$ is the coherent pump amplitude at site $n$, that oscillates in time $t$ with a frequency $\omega_{\rm P}$ which we consider to be the same for all $n$; and finally, $U$ is a Kerr-type non-linear frequency shift which can appear, e.g., in polaritonic lattices due their matter component~\cite{CarusottoCiuti2013}, and that we also assume to be equal at all sites.

To find the steady states of Eqs.~\eqref{eq:coupled mode} we use a 4th order Runge-Kutta routine to solve the dynamics and run it until converged results are obtained. Codes to reproduce the figures of the paper are available in~\cite{GitHub}.

\section{Localization in a one-dimensional photonic lattice within the linear regime}
\label{sec:linear}

Let us start by reviewing and extending the results obtained in Refs.~\cite{Jamadi2022,Gonzalez-Tudela_2022} in the linear regime for one-dimensional photonic lattices [i.e., taking $U=0$ in Eq.~\eqref{eq:coupled mode}]. To analyze the emergence of driven-dissipative localization in such scenario, it is convenient to choose a rotating frame with the laser frequency $\omega_{\rm P}$. Therefore, one can write the field amplitudes in Eq.~\eqref{eq:coupled mode} as $a_n\rightarrow a_n e^{-i\omega_{\rm P}t}$. Then, it is also convenient to assume periodic boundary conditions to expand $a_n$ in its Fourier components, i.e., $a_n=(1/\sqrt{N})\sum_k a_k e^{-i k n}$, with $k$ being the associated momenta. Putting together these two considerations, Eq.~\eqref{eq:coupled mode} can be written as
\begin{align}
    i\dot{a}_k = [\omega(k)-\omega_{\rm P}] a_k
    -i\gamma a_k
    +F_k,
\label{eq:coupled-modek1}
\end{align}
with $F_k = (1/\sqrt{N})\sum_n F_n e^{-ikn}$ being the $k$-dependence of the coherent pump, and $\omega(k)$ being the energy dispersion of the photonic lattice which reads
\begin{align}
\omega(k)=\omega_a-2J\cos(k).
\label{eq:endisp1d}
\end{align}
The steady state solution of Eq.~\eqref{eq:coupled-modek1} can be found by setting $\dot{a}_k=0$, obtaining
\begin{align}
a_k(t\rightarrow\infty)=\frac{F_k}{\omega_{\rm P}-\omega(k)+i\gamma}.
\end{align}
In the equation above, we can see that the most populated modes in momentum space will be those satisfying
\begin{align}
    \omega(k_a)=\omega_{\rm P}\rightarrow k_a=\pm \arccos\left(\frac{-\Delta}{2J}\right),
\label{eq:resonance}
\end{align}
where $\Delta = \omega_{\rm P} - \omega_a$ is the pump-resonator detuning. The predominance of such values of momentum makes it possible that, when several pumps drive the photonic lattice at different sites, they can interfere constructively or destructively depending on their relative position. For instance, if one drives the lattice at two positions $n_{1,2}$, the emission from each spot will acquire a phase $e^{ik_a d}$ after travelling through the inter-pump distance $d=n_1-n_2$. Thus, if the drivings are equal in amplitude and phase, when the distance $d$ and momentum $k_a$ are such that
\begin{align}
    1+e^{ik_{a}d} = 0,
\label{eq:localization}
\end{align}
destructive interference occurs at all sites to the left and to the right of both spots, and therefore the emission remains fully localized in between the two drivings. This mechanism is schematicaly depicted in Fig.~\ref{fig:scheme}(a). For instance, at a detuning $\Delta=0$ where the resonant pump momenta given by Eq.~\eqref{eq:resonance} are $k_a = \pm\pi/2$, localization occurs at distances $d=2(2\ell+1)$, where $\ell\in\mathbb{Z}$. As shown in Refs.~\cite{Jamadi2022,Gonzalez-Tudela_2022}, localization can be perfect in the limit of small losses, i.e., when $\gamma\rightarrow 0$.

In Fig.~\ref{fig:linear} we illustrate this localization for two different distances between the pumping spots, $d=2$ and $d=6$, in the left and middle column, respectively. To quantify localization, we define the following parameter:
\begin{align}
    \lambda = \frac{\sum_{n_1<n<n_2} |a_n|^2}{I_{\rm T}},
\end{align}
which accounts for the fraction of intensity in the region between the pumps with respect to the total intensity inside the lattice $I_{\rm T}=\sum_n|a_n|^2$. In Fig.~\ref{fig:linear}(a,c) we plot as blue circles the localization parameter $\lambda$ and the total intensity [panels (b,d)] as a function of the detuning $\Delta$ for $d=2,6$, obtained by solving numerically the coupled-mode Eqs.~\eqref{eq:coupled-modek1}. There, one can observe that for $d=2$ [panel (a)] $\lambda$ has a single maximum (at $\Delta=0$), whereas for $d=6$ [panel (c)] several maxima appear. This can be understood by substituting the value of the resonant momentum $k_{\rm a}$ given by Eq.~\eqref{eq:resonance} into the localization condition in Eq.~\eqref{eq:localization}. This implies that the values of detuning at which localization maxima appear are given by 
\begin{align}
    \Delta_\ell = -2J\cos\left[\frac{(2\ell+1)\pi}{d}\right],
\label{eq:Delta localization}
\end{align}
where $\ell\in\mathbb{Z}$. For $d=2$ the equation above only allows a localization maximum at $\Delta=0$. However, for $d=6$ there are three integers ($\ell=0,1,2$) leading to localization maxima at three different values $\Delta\simeq -1.73J$, $0$, and $1.73J$.

Both for $d=2$ and $d=6$, when one takes a value of $\Delta$ leading to a localization maximum, the steady-state spatial population is localized between the pumping spots. As an example, Fig.~\ref{fig:linear}(e,f) shows respectively the intensity distribution in real space $|a_n|^2$ with an inter-pump distance $d=2,6$, both for a fixed detuning $\Delta=0$ in which $\lambda$ features a maximum for both distances. Interestingly, these localization maxima appear in frequency regions where the total intensity $I_{\rm T}$ is minimum, as shown in Fig.~\ref{fig:linear}(b,d). This has passed inadvertently in previous works and will have important consequences in the appearance of non-linear effects, as we will see in Sec.~\ref{sec:1d non-linear}.

To gain insight on the interplay between localization and total intensity, we obtain approximate analytical expressions for both the spatial profiles of the populations, $|a_n|^2$, as well as the total intensity $I_{\rm T}$, which read (see Supplement 1 for details)
\begin{align}
    |a_n|^2= \left\{ \begin{array}{l} 2\pi^2|F|^2 D(\Delta)^2 e^{-\pi\gamma D(\Delta) (n_1+n_2-2n)}
    \\
    \;\;\;\;\;\;
    \times\left[ \cosh{\left(\pi\gamma D(\Delta)d\right)} + \cos{\left(k_0 d\right)} \right] \;\; \text{if} \;\; n \leq n_1 \\ 
    2\pi^2|F|^2 D(\Delta)^2 e^{-\pi\gamma D(\Delta)d}
    \\
    \;\;\;\;\;\;
    \times    
    [ \cosh{\left(\pi\gamma D(\Delta)(n_1+n_2-2n)\right)}
    \\
    \;\;\;\;\;\;
    +\cos{\left(k_0 (n_1+n_2-2n)\right)} ]\;\; \text{if} \;\;  n_1 \leq n \leq n_2\\
    2\pi^2|F|^2 D(\Delta)^2 e^{\pi\gamma D(\Delta) (n_1+n_2-2n)}
    \\
    \;\;\;\;\;\;
    \times
    \left[ \cosh{\left(\pi\gamma D(\Delta)d\right)} + \cos{\left(k_0 d\right)} \right]\;\; \text{if} \;\;    n \geq n_2\end{array} \right. ,
\label{eq:analytic intensity}
\end{align}
and
\begin{align}
    I_{\rm T} = \frac{2\pi|F|^2 D(\Delta)}{\gamma}\left[ 1 + \cos{(k_0 d)} e^{-\pi\gamma D(\Delta)d} \right]\,,
\label{eq:analytic total intensity}
\end{align}
respectively, where $D(\Delta)$ is the density of states of the one-dimensional photonic lattice:
\begin{align}
    D(\Delta) = \frac{1}{N}\sum_k \delta(\Delta+2J\cos{k}) = \frac{1}{\pi}\frac{1}{\sqrt{4J^2-\Delta^2}}\,.
\end{align}
To our knowledge, the above expressions have never been obtained before and provide a very clear picture of the dependence of the results with some relevant photonic lattice figures of merit, like the density of states or photon-decay rate. By plotting them in the dashed lines in Fig.~\ref{fig:linear}(a-d) we evidence that they feature a perfect agreement with the numerical results. The only significant deviations appear near the band-edges, $\Delta=\pm 2 J$, where the approximations used to derive Eqs.~(\ref{eq:analytic intensity}-\ref{eq:analytic total intensity}) are not valid (see Supplement 1 for details).

From the analytical expression of the total intensity, Eq.~\eqref{eq:analytic total intensity}, we can obtain two conclusions. First, we can immediately see that the localization condition of Eq.~\eqref{eq:localization} implies at the same time a diminished total intensity. In Supplement 1, we explain that in terms of the analogy put forward in Ref.~\cite{Gonzalez-Tudela_2022}, and show that $I_{\rm T}$ is just proportional to the local density of states (LDOS)~\cite{Gonzalez-Tudela2017prl,Gonzalez-Tudela2017pra,Barnes2020} of two emitters coupled to the bath at sites $n_1$ and $n_2$. Thus, this decrease of the total intensity can be understood as an inefficient coupling of the ensemble of the two pumping spots to the photonic lattice.
Second, the total intensity is maximized due to two effects: by the increase of the density of states $D(\Delta)$ in slow light regions (i.e., at the band-edges of our 1D lattice, where the group velocity $v_{\rm g}=\partial\omega/\partial k=[\pi D(\Delta)]^{-1}$ vanishes~\cite{Baba2008,Calajo2016}), and, complementary, at the regions corresponding to constructive interference, i.e., $1-e^{ik_a d}=0$ for this choice of pumping spot phases, where there is a collective enhancement of the intensity.

This condition of constructive interference allows us to calculate a similar expression to Eq.~\eqref{eq:Delta localization}, but in this case accounting for the values of detuning where $I_{\rm T}$ features a maximum:
\begin{align}
    \Delta_\ell = -2J\cos\left[\frac{2\ell\pi}{d}\right],
\label{eq:Delta maximum I_T}
\end{align}
where $\ell\in\mathbb{Z}$. This is in agreement with the results of our numerical simulations shown in Fig.~\ref{fig:linear}(b,d), where maxima of $I_{\rm T}$ can be found at $\Delta=-2J$ and $2J$ for $d=2$, and at $\Delta=-2J$, $-J$, $J$, and $2J$ for $d=6$. These values of $\Delta$ also correspond to minima of $\lambda$.

Finally, apart from explaining such an interplay between localization and total intensity, Eqs.~(\ref{eq:analytic intensity}-\ref{eq:analytic total intensity}) allow understanding the dependence of both $\lambda$ and $I_{\rm T}$ for different inter-pump distances $d$. Since both $d$ and $\gamma$ appear in the exponential function of Eq.~\eqref{eq:analytic total intensity}, we expect that increasing one of such quantities will spoil perfect localization, and conversely produce an increase in $I_{\rm T}$. This is indeed observed in Fig.~\ref{fig:linear}(g-h): for a fixed detuning $\Delta=0$ the localization (total intensity) decreases (increases) for growing $d$. Such results are confirmed by the numerical simulations shown as blue circles.

\section{Non-linear driven-dissipative localization effects: stability and enhancement}
\label{sec:1d non-linear}

In this Section, we study how the driven-dissipative localization scenario changes when one includes sizeable Kerr non-linearities, i.e., $U\neq 0$. Without loss of generality, we restrict our study to positive values of $U$. As can be seen in Eq.~\eqref{eq:coupled mode}, we consider local non-linearities that modify the on-site energy of each site by a factor $U|a_n|^2$. A priori, since the steady-state population profiles $|a_n|^2$ are not uniform [see Fig.~\ref{fig:linear}(e,f)], it is unclear whether localization survives in the non-linear regime, and if it does, whether it enhances non-linear phenomena such as optical bistabilities. We answer these two questions in Sections~\ref{subsec:stability} and ~\ref{subsec:bistability}, respectively.

\subsection{Stability of driven-dissipative localization~\label{subsec:stability}}

\begin{figure}[tb]
    \centering
    \includegraphics[width=\linewidth]{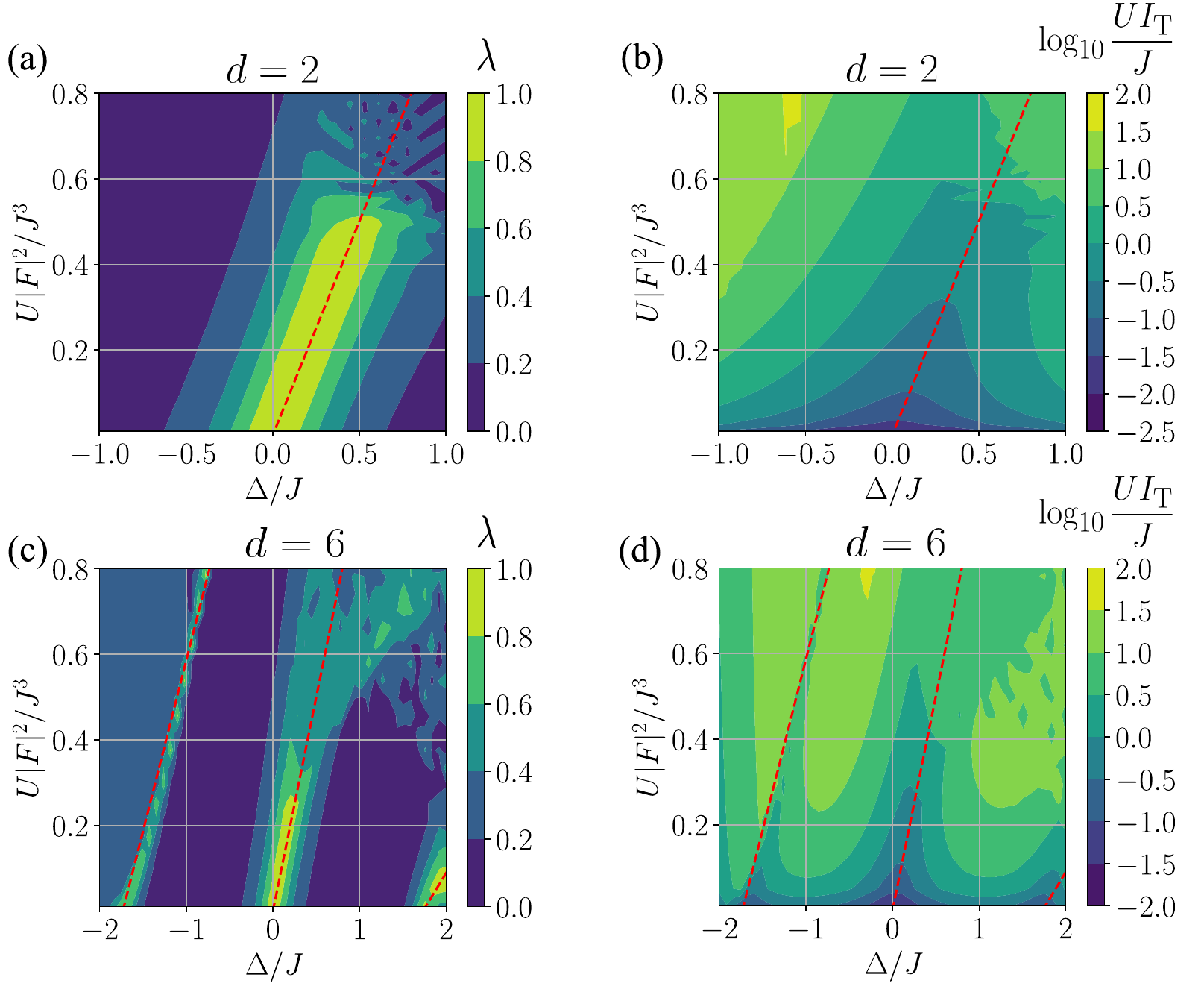}
    \caption{Localization in the non-linear 1D lattice. (a) [(c)] Localization parameter $\lambda$ for an inter-pump distance $d=2$ [$d=6$] as a function of the pump amplitude $|F|^2$ (normalized in units of the Kerr non-linearity strength $U$ and the tunneling rate $J$) and the detuning $\Delta$. (b) [(d)] Analog plot for the total intensity inside the lattice $I_{\rm T}$, calculated for $d=2$ [$d=6$]. In all panels, red dashed lines signal the dependence $\Delta=\Delta_\ell+\alpha_\ell U|F|^2/J^2$ of the detuning at which localization maxima (and $I_{\rm T}$ minima) appear (see the text). Simulation parameters: cavity decay rate $\gamma = 5 \times 10^{-2} J$, $N=500$ sites with open boundary conditions, $n_1=250$.  }
    \label{fig:non-linear}
\end{figure}

Let us start by analyzing the behavior of the localization parameter $\lambda$ and the total intensity $I_{\rm T}$ for a growing non-linearity. For that, we produce contour plots of $\lambda$ and $I_{\rm T}$ as function of the adimensional quantities $\Delta/J$ and $U|F|^2/J^3$ in Fig.~\ref{fig:non-linear}(a,c) and (b,d), respectively. Panels (a,b) are calculated for an inter-pump distance $d=2$, while panels (c,d) correspond to $d=6$.

Focusing first on the localization parameter, we observe that, contrary to intuition, localization survives for non-zero values of $U$ in frequency regions forbidden in the linear regime. Let us consider, for example, the case of $d=2$, where localization in the linear regime occurs only at $\Delta=0$. In the non-linear regime, however, the value of $\Delta$ leading to the maximum localization follows the relation $\Delta=U |F|^2/J^2$ [signalized by the red dashed line in Fig.~\ref{fig:non-linear}(a,b)]  
up to a point where it suddenly decreases due to dynamical instabilities induced by the parametric couplings introduced by the Kerr non-linearity~\cite{Shukla1986,Nakazawa1988,Trillo1991,Carusotto2004,Carusotto2013} (more information about this phenomenon can be found in Supplement 1). Nevertheless, the driven-dissipative localization remains very high (i.e, $\lambda\geq 0.9$) for values of detuning approximately in the region $\Delta\in (0,0.5J)$, something which would be impossible within the linear regime. We label this phenomenon \textit{non-linear-enabled localization}.

In Fig.~\ref{fig:non-linear}(c-d) a similar analysis is carried for an inter-pump distance $d=6$. In agreement with the results of Sec.~\ref{sec:linear}, we observe that in the linear regime (i.e., for $U=0$) the localization features maxima for three detunings given by Eq.~\eqref{eq:Delta localization}, which are shifted towards higher values of $\Delta$ when $U\neq 0$, following the relation $\Delta = \Delta_\ell + \alpha_\ell U|F|^2/J^2$, with $\alpha_\ell$ being a constant that depends on $\ell$. While for the central localization region given by $\Delta_1$ we can take $\alpha_1=1$ (thus recovering the dependence of the $d=2$ case), $\alpha_\ell$ is larger than $1$ for the other two localization regions appearing for $\Delta_0$ (which is fitted by $\alpha_0\simeq 1.2$) and $\Delta_2$ (for which we estimate $\alpha_2\simeq 2.8$). This is evident from Fig.~\ref{fig:non-linear}(c), where we plot $\Delta=\Delta_\ell + \alpha_\ell U|F|^2/J^2$ as red dashed lines for $\ell=0,1,2$. Furthermore, another difference with respect to $d=2$ is that the dynamical instabilities appear for smaller values of $\Delta$ than in the $d=2$ case. 

We now focus on the behavior of the total intensity. In both cases, $d=2$ and $d=6$, the non-linear regime inherits the inverse relation between the localization parameter $\lambda$ and the total intensity $I_{\rm T}$ found in Sec.~\ref{sec:linear} for the linear case. This is evidenced by Fig.~\ref{fig:non-linear}(b,d), which displays analog contour plots of the  total intensity $I_{\rm T}$ as a function of $\Delta/J$ and $U|F|^2/J^3$ for $d=2$ [panel (b)] and $d=6$ [panel (d)]. Comparing these results with those of panels (a,c) for the localization parameter, we see that, also in the non-linear regime, a large value of $\lambda$ is accompanied by a small $I_{\rm T}$. To further clarify this point, the red dashed lines signaling the regions of maximum $\lambda$ in panels (a,c) were also plotted in panels (b,d), demonstrating that these regions correspond to minima of $I_{\rm T}$.

\begin{figure}[tb]
    \centering
    \includegraphics[width=\linewidth]{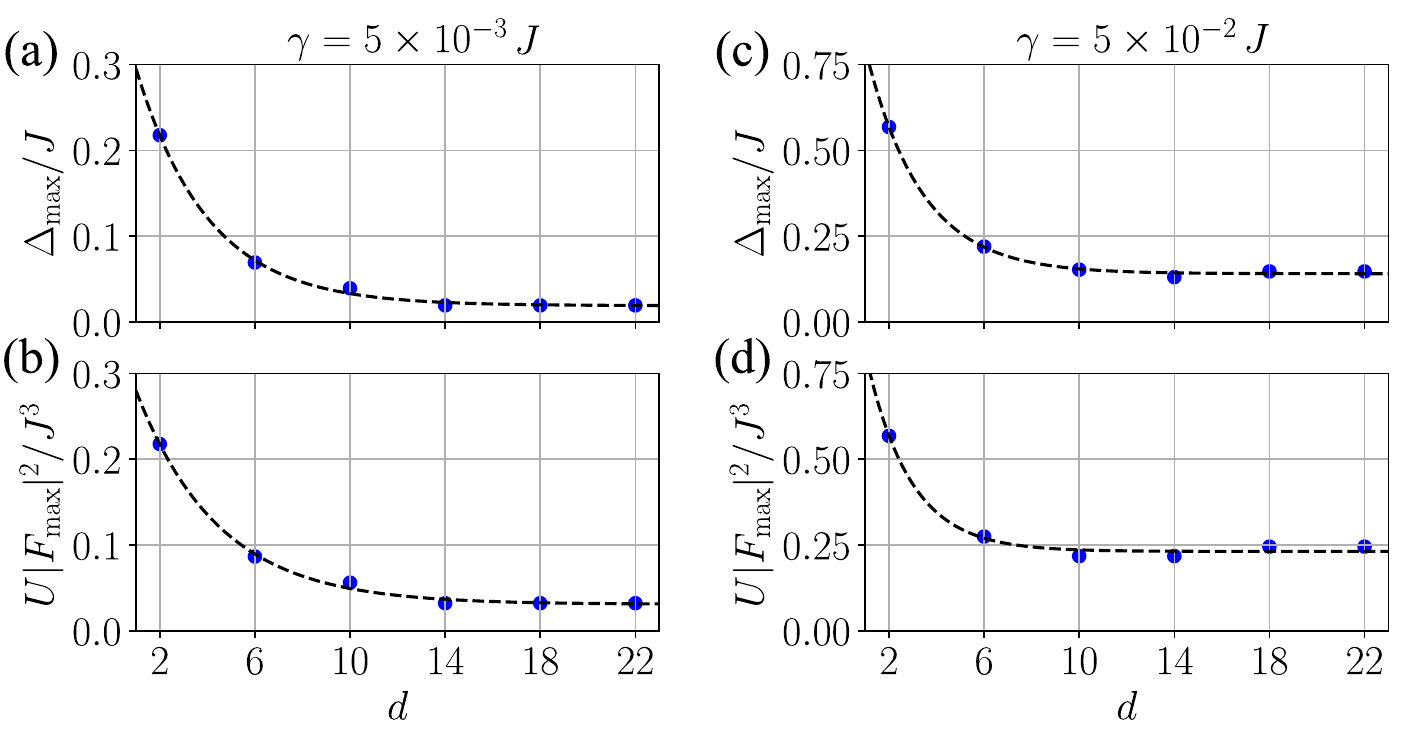}
    \caption{(a,b) Maximum values of detuning $\Delta_{\rm max}$ (normalized in units of the tunneling rate $J$) and pump intensity $|F_{\rm max}|^2$ (normalized in units of the Kerr non-linearity strength $U$ and $J$) as a function of the inter-pump distance $d$, for the central localization region with detuning $\Delta=0$ at $F=0$. The cavity decay rate is fixed at $\gamma=5\times 10^{-3}J$. Blue dots are numerical results, while dashed lines correspond to exponential fits (see the text). (c,d) Analog plots for $\gamma=5\times 10^{-2}J$. Simulation parameters: $N=500$ sites with open boundary conditions, $n_1=250$.}
    \label{fig:larger d}
\end{figure}

Finally, to analyze in more detail the limits of the non-linear enabled localization, in Fig.~\ref{fig:larger d} we plot the maximum values of $\Delta$ and $|F|^2$ in which localization drops below $\lambda = 0.9$ as a function of $d$, considering always the central localization region that appears at $\Delta=0$ when $U=0$. We label such maximum values as $\Delta_{\rm max}$ and $|F_{\rm max}|^2$, respectively. Panels (a,b) are calculated for a cavity decay rate $\gamma=5\times 10^{-3}J$, while (c,d) show analog plots for $\gamma=5\times 10^{-2}J$. For the two values of $\gamma$, we observe that both $\Delta_{\rm max}$ and $|F_{\rm max}|^2$ follow an exponential decay and then saturate, reaching constant values which are independent on $d$. This is signalized by the dashed lines, which correspond to a fit to an exponential law, $f(x)=ae^{-bx}+c$, where $a,b,c$ are the fit parameters. However, this saturation occurs for a smaller value of $d$ (around $d \simeq 10$) for $\gamma=5\times 10^{-2}J$, while for $\gamma=5\times 10^{-3}J$ it takes place around $d \simeq 14$. Moreover, for the same $d$, the values of $\Delta_{\rm max}$ and $|F_{\rm max}|^2$ are larger in the former case. This can be related to the role played by the cavity decay rate in stabilyzing the steady-state solution of the coupled-mode equations~\eqref{eq:coupled mode}.

\subsection{Enhancement of optical bistabilities~\label{subsec:bistability}}

\begin{figure*}[tb]
    \centering
    \includegraphics[width=0.7\linewidth]{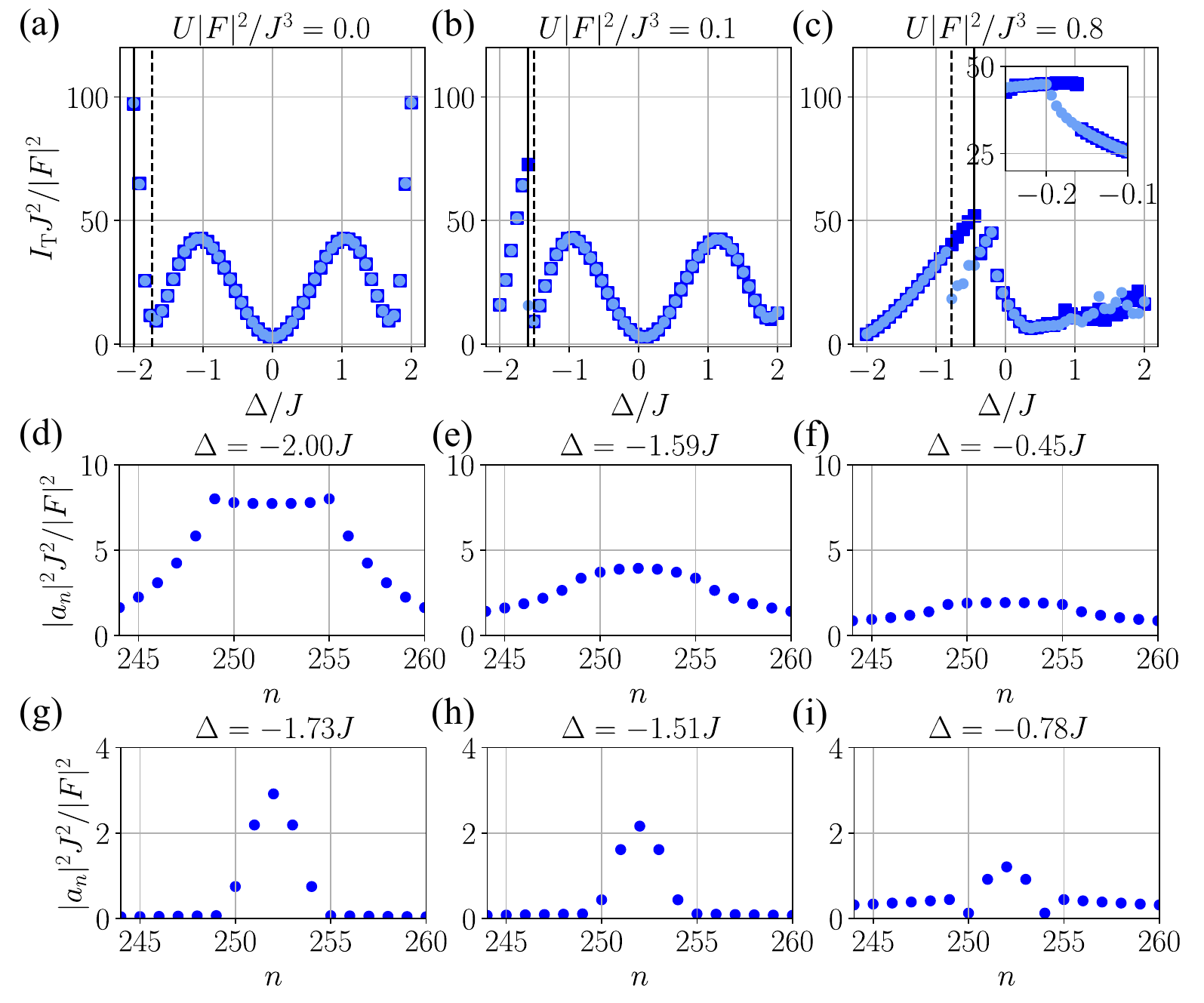}
    \caption{(a-c) Total intensity inside the lattice $I_{\rm T}$ normalized in units of the tunneling rate $J$ and the pump intensity $|F|^2$ as a function of the detuning $\Delta$, for an inter-pump distance $d=6$. Each panel corresponds to a different value of the Kerr non-linearity strength $U$. To make optical bistabilities visible, squares are the results of a rightwards ramp (in which $\Delta$ is increased starting from the lower band edge), while circles belong to a leftwards ramp (in which $\Delta$ is decreased starting from the upper band edge). (d-f) [(g-i)] Intensity $|a_n|^2$ in each site $n$ of the lattice for values of $\Delta$ in which $I_{\rm T}$ reaches the maximum [minimum] signalized by the vertical solid [dashed] lines in panels (a-c). Panels (d,g), (e,h), and (f,i) are calculated for $U|F|^2/J^3=0$, $U|F|^2/J^3=0.1$, and $U|F|^2/J^3=0.8$, respectively. Simulation parameters: cavity decay rate $\gamma = 5 \times 10^{-2} J$, $N=500$ sites with open boundary conditions, $n_1=250$.
    }
    \label{fig:bistability}
\end{figure*}

In principle, one would expect that driven-dissipative localization enhances non-linear effects due to the concentration of light into a small number of sites, see Fig.~\ref{fig:linear}(e,f). However, the trade-off between localization and total intensity in the linear and non-linear regimes that we demonstrate in the previous Section points to the opposite direction. In this Section, we study the emergence of optical bistable behavior as a signature of optical non-linear effects, and demonstrate that, actually, the appearance of bistabilities is linked to the enhancement of the total intensity due to slow light and constructive interference rather than localization regions.

To show that, in Fig.~\ref{fig:bistability}(a-c) we plot the bistability curves of the total intensity $I_{\rm T}$ for two pumps separated by a distance $d=6$, and increasing values of the adimensional parameter $U|F|^2/J^3$ quantifying the Kerr non-linearity strength. The cyan circles (blue squares) represent the intensity for a rightwards (leftwards) ramp, i.e., for $\Delta$ going from negative (positive) to positive (negative) values. In panel (a), we plot the situation with $U=0$, i.e., the linear regime. In this case, there are obviously no bistabilities, and thus squares and circles coincide. We accompany this figure by panels (d) and (e), where we plot the spatial patterns at $\Delta=-2J$ and $-1.73J$, corresponding to the first maximum and minimum of $I_{\rm T}$ in panel (a) (starting from the bottom band edge), and indicated with black solid and dashed lines, respectively. There, we see that the minimum (maximum) intensity corresponds to the situation with (no) localization.

We now study the displacement of such maximum and minimum of $I_{\rm T}$ for growing $U|F|^2/J^3$. In Fig.~\ref{fig:bistability}(b), we plot the total intensity for a small value of  $U|F|^2/J^3=0.1$. In this case, the Kerr-non-linearity blue-shifts the maxima and minima of $I_{\rm T}$ towards larger values of detuning. In particular, the slow light enhancement of $I_{\rm T}$, which takes place at $\Delta=-2J$ in the linear case, now generates a single bistability curve around $\Delta\simeq -1.59J$. It makes sense that such maximum is the first one developing an optical bistability for growing $U|F|^2/J^3$, as it is the one reaching the largest $I_{\rm T}$. On the other hand, the first minimum of $I_{\rm T}$, located at $\Delta=-1.73J$ for $U=0$, is now displaced to $\Delta=-1.51J$. In panels (e) and (h) we plot the spatial patterns corresponding to such maximum (mininum) indicated by the solid (dashed) line in panel (b), showing that when $I_{\rm T}$ reaches its maximum light spreads over all lattice sites, while when $I_{\rm T}$ is minimum light is localized in the region between the pumps.

\begin{figure*}[tb]
    \centering
    \includegraphics[width=\linewidth]{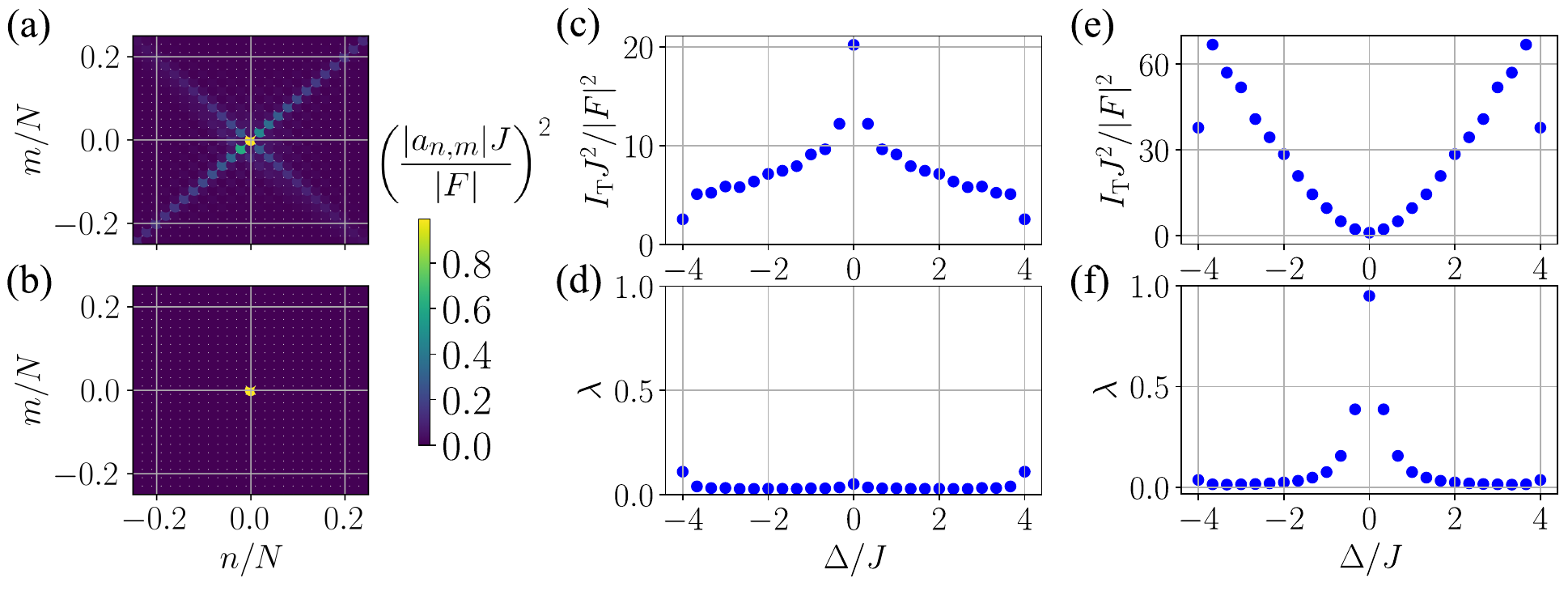}
    \caption{Localization in the linear ($U=0$) square lattice. (a) Spatial distribution of the intensity $|a_{n,m}|^2$ (normalized using the tunneling rate $J$ and the pump intensity $|F|^2$) for an $N\times N$ lattice coherently driven at its central site $(0,0)$ (i.e., \textit{individual driving}). The detuning is fixed at $\Delta=0$. (b) An analog plot for a lattice driven by four pumps located at $(\pm 1,0)$ and $(0, \pm 1)$ (i.e., \textit{collective driving}), where the inter-site distance is $1$. (c) Total intensity in the lattice $I_{\rm T}$ in the individual driving configuration as a function of the detuning $\Delta$. (d) Analog plot for the localization parameter $\lambda$.
    (e) Total intensity in the lattice $I_{\rm T}$ in the collective driving configuration as a function of the detuning $\Delta$. (f) Analog plot for the localization parameter $\lambda$.
    Simulation parameters: cavity decay rate $\gamma = 5 \times 10^{-2} J$, $N=50$ (i.e, $50\times 50$ sites) with open boundary conditions.
    }
    \label{fig:2d linear}
\end{figure*}

This behavior continues up to a critical value of $U|F|^2/J^3$ in which a second hysteresis cycle opens at the second intensity maximum (located at $\Delta=-J$ in the linear regime, and blue-shifted for a finite $U|F|^2/J^3$). Fig.~\ref{fig:bistability}(c) shows the total intensity for $U|F|^2/J^3=0.8$, for which both optical bistabilities can be observed (the second one is displayed in an inset). At $\Delta \gtrsim 0$ a dynamical instability arises, in agreement with Fig.~\ref{fig:non-linear}(b). Such a large value of $U|F|^2/J^3$ further blue-shifts the first maximum (minimum) of $I_{\rm T}$, which now take place at $\Delta = -0.45J$ ($\Delta=-0.78J$), and is signalized by the solid (dashed) line. The spatial distribution of the intensity for these values of $\Delta$ is plotted in Fig.~\ref{fig:bistability}(f,i). The results of these panels confirm that a large intensity inside the lattice corresponds to light spreading over several resonators, while localization is found at values of detuning in which $I_{\rm T}$ is minimum. Actually, in panel (i) localization is not perfect, as intensity leaks outside of the region between the pumps. This can be related to the fact that for $U|F|^2/J^3=0.8$ there is a larger value of $I_{\rm T}$ at $\Delta=-0.78J$ in the lower bistability branch [see panel (c)] compared, for instance, with the smaller value of $I_{\rm T}$ that appears for $U|F|^2/J^3=0.1$ at $\Delta=-1.51J$ [see panel (b)]. This implies a better localization in the latter case, with almost no intensity outside the region between the pumps [see panel (h)].

Overall, the results of this Section demonstrate that localization is not linked to an enhancement of non-linear effects. Actually, it takes place at values of detuning for which the total intensity $I_{\rm T}$ features a local minimum. On the contrary, non-linear phenomena such as optical bistabilities appear when the intensity inside the lattice is maximum. Such situation is associated to either slow light effects or a constructive interference between the two pumps, implying a spreading of light over the whole lattice.

\section{Extension to two-dimensional lattices}
\label{sec:2d}

In this Section we show that the previous conclusions of the non-linear driven-dissipative localization can also be extended to 2D lattices. In particular, we consider a square lattice formed by $N\times N$ sites with first-neighbor couplings [depicted in Fig.~\ref{fig:scheme}(b)].  We start by addressing the linear regime. This was extensively studied in Refs.~\cite{Jamadi2022,Gonzalez-Tudela_2022}, where it was shown that the driven-dissipative localization can also occur in 2D geometries. However, this is again limited to specific frequencies, as it was for 1D lattices.

As a first step, we reproduce the behavior of linear square lattices observed in Ref.~\cite{Gonzalez-Tudela_2022}, as well as check that the inverse relation between localization and total intensity unveiled in the 1D case also holds in 2D. By setting $U=0$, the coupled-mode equations~\eqref{eq:coupled mode} for a square lattice in reciprocal space and in the reference frame rotating at the frequency of the coherent pumps can be written as
\begin{align}
    i\dot{a}_{k_x,k_y} = [\omega(k_x,k_y)-\omega_{\rm P}] a_{k_x,k_y}
    -i\gamma a_{k_x,k_y}
    +F_{k_x,k_y},
\label{eq:coupled mode k square}
\end{align}
where $k_x$ and $k_y$ are the momentum components in the $x$ and $y$ directions, $F_{k_x,k_y} = (1/N)\sum_{n,m}$ $F_{n,m} e^{-i(k_x n+k_y m)}$ is the Fourier transform of the coherent drive, $F_{n,m}$ is the pump amplitude at site $(n,m)$ (where the indices $n,m=-N/2,\dots,N/2-1$ label the lattice sites), and the dispersion relation is given by
\begin{align}
    \omega(k_x,k_y)=\omega_a-2J\left[\cos(k_x)+\cos(k_y)\right],
\label{eq: disperion relation square}
\end{align}
where $J$ is the tunneling amplitude. In this case, the frequency band extends from $\Delta=-4J$ to $4J$.

By solving for the steady-state of the coupled-mode equations, we can calculate the spatial distribution of the intensity inside a $50\times 50$ lattice. We choose a fixed value of detuning $\Delta=0$, i.e., at the center of the frequency band. When a single resonant pumping spot is placed at the central site $(0,0)$ (we refer to this configuration as \textit{individual driving}), light is emitted into the lattice in a collimated fashion, as depicted in Fig.~\ref{fig:2d linear}(a). This is due to the dispersion relation~\eqref{eq: disperion relation square}, which leads to a non-uniform group velocity~\cite{Gonzalez-Tudela2017pra,Gonzalez-Tudela2017prl}. As shown in~\cite{Gonzalez-Tudela_2022}, such collimated emission allows to find a perfect localization by coupling four pumps at positions $(\pm 1,0)$ and $(0, \pm 1)$ (we label this configuration \textit{collective driving}). The resulting spatial intensity distribution is shown in Fig.~\ref{fig:2d linear}(b), where one can clearly observe the localization of light in the central spot. 

 This behavior is in agreement with the results of~\cite{Gonzalez-Tudela_2022}. We now extend them by studying in detail the properties of the total intensity $I_{\rm T}$ as well as the localization parameter $\lambda$ as the detuning $\Delta$ is varied across the lattice spectrum. Both $I_{\rm T}$ and $\lambda$ feature a very different behavior in the individual and collective driving configurations. In the former case, the total intensity [see Fig.~\ref{fig:2d linear}(c)] exhibits a maximum at $\Delta=0$ (i.e., when  $\omega_\mathrm{P}=\omega_a$) due to the saddle-point region appearing at the center of the dispersion relation. This leads to a vanishing group velocity in a single direction, which is associated to a diverging density of states~\cite{Ashcroft1976}. As we learned from the 1D case, this implies a peak in $I_{\rm T}$ at the corresponding value of $\Delta$. Regarding the localization parameter $\lambda$ [see Fig.~\ref{fig:2d linear}(d)], it is negligible throughout the whole spectrum, except at its edges (i.e., at $\Delta=\pm 4J$), where a small increase is appreciated, corresponding to the vanishing $I_{\rm T}$. On the other hand, in the collective driving configuration the destructive interference giving rise to the localization shown in Fig.~\ref{fig:2d linear}(b) leads to a negligible $I_{\rm T}$ around $\Delta=0$, irrespective of the vanishing group velocity at that value of detuning [see Fig.~\ref{fig:2d linear}(e)]. As in the 1D lattice, associated to the minimum of $I_{\rm T}$, in Fig.~\ref{fig:2d linear}(f) a maximum of localization in which $\lambda\simeq 1$ appears at $\Delta=0$.

\begin{figure}[tb]
    \centering
    \includegraphics[width=\linewidth]{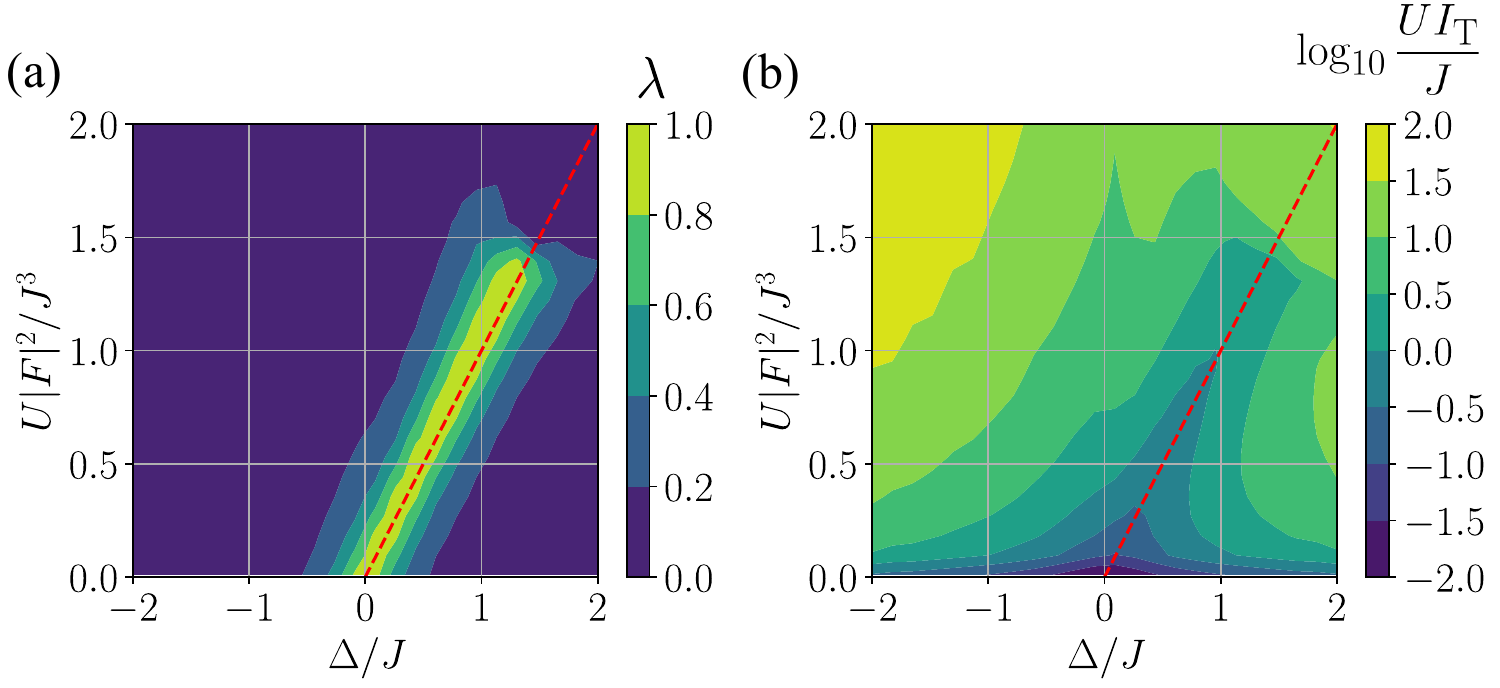}
    \caption{Localization in the non-linear square lattice. (a) Localization parameter $\lambda$ as a function of the pump amplitude $|F|^2$ (normalized in units of the Kerr non-linearity strength $U$ and the tunneling rate $J$) and the detuning $\Delta$. The red dashed line represents the linear dependence $\Delta = U|F|^2/J^2$. (b) Analog plot for the total intensity inside the lattice $I_{\rm T}$. Simulation parameters: cavity decay rate $\gamma = 5 \times 10^{-2} J$, $N=50$ (i.e., $50\times 50$ sites) with open boundary conditions.  }
    \label{fig:2d non-linear}
\end{figure}

Overall, we can conclude that in the square lattice there exists an analog trade-off between localization and total intensity to the one that we already found for the 1D lattice. This hints to a similar behaviour when non-linearities are included in the 2D case. This is what we test in Fig.~\ref{fig:2d non-linear}, where we show contour plots of the localization parameter $\lambda$ [panel (a)] as well as the total intensity $I_{\rm T}$ [panel (b)] as a function of the adimensional parameters $\Delta/J$ and $U|F|^2/J^3$. On the one hand, a finite Kerr non-linearity enables localization for a range of detunings $\Delta/J\in (0,1.35)J$, far beyond what can be found in the linear regime, in which localization is restricted to $\Delta = 0$. Above the upper limit of the non-linear case, located at $\Delta\simeq 1.35J$, dynamical instability suppresses localization, which experiences a sudden drop. As for a 1D lattice with inter-pump distance $d=2$, the region of maximum $\lambda$ follows a linear dependence $\Delta=U|F|^2/J^2$, signalized by the red dashed line in panel (a). In terms of intensity, panel (b) shows that a non-zero localization is accompanied by a decreasing $I_{\rm T}$. To evidence this, we have also plotted the $\Delta=U|F|^2/J^2$ dependence followed by the region with high localization in panel (b), showing that the area in which $I_{\rm T}$ is minimum adheres to the same behavior.

We finish this Section by noting that the same conclusions we obtained for the 1D lattice also hold for this 2D lattice: A finite non-linearity stabilizes localization for a larger range of frequencies than in the absence of non-linearities. Besides, a large localization is associated with a vanishing intensity inside the lattice, which prevents the exploitation of this phenomenon for the enhancement of non-linear effects.

\section{Conclusions and outlook}
\label{sec:discussion}

Summing up, we have studied the phenomenon of driven-dissipative localization in one and two-dimensional lattices in the presence of optical non-linearities. We find that the non-linearity allows to expand the frequency region in which localization can be obtained compared to the linear regime. Besides, contrary to our initial intuition, such localization does not enhance non-linear effects due to the concomitant decrease of global intensity inside the lattice. In fact, non-linear effects such as optical bistabilities occur in regions where the emission between the pumping spots interferes constructively, maximizing the total intensity. We foresee that our results will pave new ways of harnessing non-linear effects in driven-dissipative systems, e.g., as reconfigurable optical simulator~\cite{Berloff2017,Kalinin2018,Kalinin2019}. Another direction is to consider the interplay of such driven-dissipative non-linear effects with topologically non-trivial band structures~\cite{Ozawa2019}.

\begin{acknowledgments}
  The authors acknowledge support from the Proyecto Sin\'ergico CAM 2020 Y2020/TCS-6545 (NanoQuCo-CM), the CSIC Research Platform on Quantum Technologies PTI-001 and from Spanish projects PID2021-127968NB-I00 and TED2021-130552B-C22 funded by r MCIN/AEI/10.13039/501100011033/FEDER, UE and MCIN/AEI/10.13039/501100011033, respectively. AMH acknowledges support from Fundación General CSIC's ComFuturo programme which has received funding from the European Union's Horizon 2020 research and innovation programme under the Marie Skłodowska-Curie grant agreement No. 101034263. AA acknowledges support from European Research Council grant EmergenTopo (865151), the French government through the Programme Investissement d’Avenir (I-SITE ULNE /ANR-16-IDEX-0004 ULNE) managed by the Agence Nationale de la Recherche, the Labex CEMPI (ANR-11-LABX-0007) and the region Hauts-de-France. AGT acknowledges support from a 2022 Leonardo Grant for Researchers and Cultural Creators, and BBVA Foundation. The authors thank Carlos Navarrete-Benlloch for discussion on the solution of non-linear coupled-mode equations.
\end{acknowledgments}

\bibliographystyle{apsrev4-2}
\bibliography{Bibliography,referencesAlex}

\clearpage

\begin{widetext}

\begin{center}
\textbf{\large Supplement 1: Non-linear-enabled localization in driven-dissipative photonic lattices\\}
\end{center}
\setcounter{equation}{0}
\setcounter{figure}{0}
\setcounter{section}{0}
\makeatletter

\renewcommand{\thefigure}{SM\arabic{figure}}
\renewcommand{\thesection}{SM\arabic{section}}  
\renewcommand{\theequation}{SM\arabic{equation}}

\section{Analytic expressions for the intensity inside the 1D lattice in the linear regime}
\label{sec:analytic intensity}

In this Section we derive analytical expressions for the spatial distribution of the intensity $|a_n|^2$ as well as for the total intensity $I_{\rm T}$ in the one-dimensional (1D) lattice with nearest-neighbour tunnelings explored in the Main Text. We first consider the case with two coherent pumps (which can lead to localization), and then for completitude we study a configuration featuring a single pump (in which localization cannot be observed). In all cases, we restrict ourselves to the linear regime (i.e., we set $U=0$) and consider a lattice formed by $N$ sites.

\subsection{Two pumps}
\label{secSM:2 pumps}

Here we assume that the lattice is coherently driven at two sites $n_1$ and $n_2 > n_1$, separated by a distance $d=n_2 - n_1$.

\subsubsection{Spatial distribution of the intensity}
\label{secSM:2 pumps a_n}

We now derive an analytical expression for the spatial distribution of the intensity $|a_n|^2$. We start from the coupled-mode equations in reciprocal space for the electric field amplitudes
\begin{align}
    i\dot{a}_k = 
    -(\Delta+2J\cos{k})a_k
    -i\gamma a_k
    +F_k,
\label{eqSM:coupled mode equations k}
\end{align}
where $k$ is the lattice momentum, $a_k = (1/\sqrt{N})\sum_n a_n e^{-ikn}$ and $F_k = (1/\sqrt{N})\sum_n F_n e^{-ikn}$ are the Fourier transforms of the field amplitude and the driving, respectively, $\Delta$ is the laser-resonator detuning, $J$ is the tunneling amplitude between neighboring sites, and $\gamma$ is the cavity decay rate.
In the steady state one has that $\dot{a}_k=0$, and therefore
\begin{align}
    a_k = \frac{F_k}{\Delta+2J\cos{k}+i\gamma}
\label{eqSM:steady state 2 pumps}
\end{align}
is the steady-state solution of Eq.~\eqref{eqSM:coupled mode equations k}.
Now we explicitly write the expression of $F_k$ for two pumps located at $n_1$ and $n_2$, which takes the shape $F_k = (Fe^{-ikn_1}/\sqrt{N})(1+e^{-ikd})$. Substituting this into Eq.~\eqref{eqSM:steady state 2 pumps}, transforming to real space, and approximating the discrete sum in momenta by an integral we get
\begin{align}
    a_n = \frac{1}{\sqrt{N}}\sum_k a_k e^{ikn} \simeq \frac{F}{2\pi}\int^{+\pi}_{-\pi}dk\frac{e^{ik|n-n_1|}+e^{-ik|n-n_2|}}{\Delta+2J\cos{k}+i\gamma}.
\label{eqSM:a_n first step}
\end{align}
Above, we have assumed that $n$ belongs in the region between the pumps, i.e. $n_1<n<n_2$. Although the following steps of the derivation correspond to this case, analog calculations can be carried out for values of $n$ outside that region. The expressions for the two other situations $n<n_1$ and $n>n_2$ are provided in Eq.~(9) of the Main Text.

We continue the derivation by expanding the denominator of Eq.~\eqref{eqSM:a_n first step} around the resonant momenta $k_0=\pm\arccos(-\Delta/2J)$ up to first order in $k$, obtaining
\begin{align}
    a_n = \frac{F}{2\pi}\int^{+\pi}_{-\pi}dk\left[ \frac{e^{ik|n-n_1|}+e^{-ik|n-n_2|}}{-\sqrt{4J^2-\Delta^2}(k-k_0)+i\gamma} + \frac{e^{ik|n-n_1|}+e^{-ik|n-n_2|}}{\sqrt{4J^2-\Delta^2}(k+k_0)+i\gamma}\right].
\end{align}
Since the integrands above vanish outside the first Brillouin zone (i.e., from $-\pi$ to $+\pi$), the above integration limits can be extended from $-\infty$ to $+\infty$. Using the residue theorem of complex analysis, we finally obtain
\begin{align}
    a_n = -i\pi F D(\Delta)\left[
    e^{ik_0|n-n_1|}e^{-\pi\gamma D(\Delta)|n-n_1|}
    + e^{ik_0|n-n_2|}e^{-\pi\gamma D(\Delta)|n-n_2|}
    \right].
\label{eqSM:amplitude 2 pumps}
\end{align}
Above, $D(\Delta)=1/(\pi\sqrt{4J^2-\Delta^2})$ is the density of states. The spatial distribution of the intensity $|a_n|^2$ is now obtained by simply taking the square modulus of Eq.~\eqref{eqSM:amplitude 2 pumps}. The resulting expression is
\begin{align}
    |a_n|^2 = 2\pi^2|F|^2 D(\Delta)^2 e^{-\pi\gamma D(\Delta)d}[ \cosh{\left(\pi\gamma D(\Delta)(n_1+n_2-2n)\right)}
    +\cos{\left(k_0 (n_1+n_2-2n)\right)} ],
\label{eqSM:intensity 2 pumps}
\end{align}
which also appears in Eq.~(9) of the Main Text.

\subsubsection{Total intensity}
\label{secSM:2 pumps I_T}

We now derive an analytic expression for the total intensity $I_{\rm T}=\sum_n |a_n|^2$. Together with Eq.~\eqref{eqSM:intensity 2 pumps}, this will allow us to calculate the localization parameter $\lambda = \sum_{n_1<n<n_2}|a_n|^2/I_{\rm T}$.
We start by noting that the total intensity can be equivalently written as $I_{\rm T} = \sum_k |a_k|^2$ in reciprocal space. Thus, one has that
\begin{align}
    I_{\rm T} \simeq \frac{|F|^2}{2\pi}\int^{+\pi}_{-\pi}dk \frac{2+e^{ikd}+e^{-ikd}}{(\Delta+2J\cos{k})^2+\gamma^2},
\label{eqSM:integral 2 pumps I_T}
\end{align}
where we have employed Eq.~\eqref{eqSM:steady state 2 pumps} to account for the steady state of $a_k$ in the presence of two pumps separated by a distance $d$, and we have already transformed the discrete sum in momenta to an integral. As we did for $|a_n|^2$, we expand the denominator around the resonant momenta $k_0=\pm\arccos(-\Delta/2J)$ up to first order in $k$, and we extend the integral from $-\infty$ to $+\infty$, obtaining
\begin{align}
    I_{\rm T} \simeq \frac{|F|^2}{2\pi}\int^{+\infty}_{-\infty}dk
    \left[
    \frac{2+e^{ikd}+e^{-ikd}}{(4J^2-\Delta^2)(k+k_0)+\gamma^2}
    + \frac{2+e^{ikd}+e^{-ikd}}{(4J^2-\Delta^2)(k-k_0)+\gamma^2}
    \right].
\end{align}
The integral above can be calculated by applying the residue theorem of complex analysis, which finally yields
\begin{align}
    I_{\rm T} = \frac{2\pi|F|^2 D(\Delta)}{\gamma}\left[
    1+\cos(k_0 d)e^{-\pi\gamma D(\Delta)d}
    \right].
\label{eqSM:2 pumps I_T}
\end{align}
This is Eq.~(10) of the Main Text.

\subsection{Single pump}
\label{secSM:1 pump}

Here we consider a single coherent pump driving the site $n_1$ of the linear ($U=0$) 1D lattice with first neighbor couplings. Below these lines we provide analytic expressions for the spatial distribution of the intensity $|a_n|^2$ as well as for the total intensity $I_{\rm T}$. The calculations are analog to those of Sec.~\ref{secSM:2 pumps}, and therefore we skip some details.

\subsubsection{Spatial distribution of the intensity}
\label{secSM:1 pump a_n}

We start with the derivation of the spatial distribution of the intensity $|a_n|^2$. As in Sec.~\ref{secSM:2 pumps a_n}, we use Eq.~\eqref{eqSM:steady state 2 pumps} to account for the steady state of the coupled-mode equations~\eqref{eqSM:coupled mode equations k}. However, for a single pump we use $F_k = (1/\sqrt{N}) F e^{-ik n_1}$ as the explicit expression for the coherent drive in Fourier space. On the following, we assume $n>n_1$. In spite of this, a completely analog calculation can be carried for $n_1<n$. Overall, we get
\begin{align}
    a_n = \frac{1}{\sqrt{N}}\sum_k a_k e^{ikn} \simeq -\frac{F D(\Delta)}{2}
    \int^{+\infty}_{-\infty}dk
    \left[
    \frac{e^{ik|n-n_1|}}{k-k_0-i\pi\gamma D(\Delta)}
    - \frac{e^{ik|n-n_1|}}{k+k_0+i\pi\gamma D(\Delta)}
    \right],
\end{align}
where we have approximated the discrete sum by an integral, whose limits can be extended to $\pm\infty$. Above, we have also expanded the denominator of the steady-state solution $\Delta+2J\cos{k}+i\gamma$ around the resonant momenta $k_0=\pm\arccos(-\Delta/2J)$ up to first order in $k$.

As a final step, we evaluate the integrals above using the residue theorem of complex analysis. This gives
\begin{align}
    a_n = -i\pi F D(\Delta)e^{ik_0|n-n_1|}e^{-\pi\gamma D(\Delta)|n-n_1|}.
\end{align}
The intensity distribution $|a_n|^2$ is simply the square modulus of the equation above. By replicating this calculation with $n_1<n$ we arrive to the expression
\begin{align}
    |a_n|^2= \left\{ \begin{array}{lcc} \pi^2|F|^2 D(\Delta)^2 e^{2\pi\gamma D(\Delta) |n-n_1|} & \text{if} & n \leq n_1 \\ 
    \pi^2|F|^2 D(\Delta)^2 e^{-2\pi\gamma D(\Delta) |n-n_1|} & \text{if} &    n \geq n_1\end{array} \right. .
\label{eqSM:1 pump a_n}
\end{align}
%

\subsubsection{Total intensity}
\label{secSM:1 pump I_T}

Finally, the total intensity $I_{\rm T}$ when a single pump is coherently driving the site $n_1$ of the lattice is given by
\begin{align}
    I_{\rm T} &= \sum_k |a_k|^2 \simeq \frac{|F|^2}{2\pi}\int^{+\pi}_{-\pi}dk
    \frac{1}{(\Delta+2J\cos{k})^2+\gamma^2}
    \nonumber\\&\simeq
    \frac{|F|^2}{2\pi}\int^{+\infty}_{-\infty}dk\left[
    \frac{1}{(4J^2-\Delta^2)(k+k_0)^2+(\pi\gamma D(\Delta))^2}
    +
    \frac{1}{(4J^2-\Delta^2)(k-k_0)^2+(\pi\gamma D(\Delta))^2}
    \right],
\end{align}
where we have employed the same approximations as in Sec.~\ref{secSM:1 pump a_n}. After computing the integrals by means of the residue theorem of complex analysis, the equation above yields
\begin{align}
    I_{\rm T} = \frac{\pi |F|^2 D(\Delta)}{\gamma}.
\label{eqSM:1 pump I_T}
\end{align}
%

\section{Analogy between multiple drivings and spontaneous emission of multiple emitters}
\label{secSM:emitter driving analogy}

Following the path opened by Ref.~\cite{Gonzalez-Tudela_2022}, in this Section we provide more details on the analogy between probing the properties of lossy photonic lattices by coherently driving them, and by means of quantum emitters. In particular, we notice that the total intensity $I_{\rm T}$ inside a driven-dissipative lattice is proportional to the local density of states (LDOS) when the pumps are replaced by quantum emitters at the same spots. The LDOS measures the number of electromagnetic modes available at a given point in space, and therefore contains information about how the radiative properties of the emitter are modified by its coupling to a structured bath~\cite{Gonzalez-Tudela2017prl,Gonzalez-Tudela2017pra,Barnes2020}. 

Below, we calculate the LDOS for the 1D and 2D lattices studied in the Main Text. We consider quantum emitters coupled to the same spots in which coherent drives would be able to produce localization. In the two cases, we restrict ourselves to the linear regime (i.e., $U=0$).

\subsection{1D lattice}
\label{secSM:LDOS 1D}

Here we consider a 1D photonic lattice formed by $N$ sites with first neighbors couplings. This is coupled to two quantum emitters at sites $n_1$ and $n_2$, separated by a distance $d=n_2-n_1$. To calculate the LDOS we first need to compute the self energy of the quantum emitter $\Sigma_{\rm e}(\Delta+i\gamma)$ for an emitter-resonator detuning $\Delta$ and a photonic cavity decay rate $\gamma$. Such a quantity captures the effect of the coupling to the lattice on the quantum emitter. If we assume an emitter-photon coupling of strength $g$, the self energy reads
\begin{align}
    \Sigma_{\rm e}(\Delta+i\gamma)
    = \frac{1}{N}\sum_k
    \frac{|ge^{-ikn_1}+ge^{-ikn_2}|^2}{\Delta+2J\cos{k}+i\gamma}.
\end{align}
The LDOS is then given by twice its imaginary part, i.e.,
\begin{align}
    \Gamma_{\rm e} = -2\rm{Im}\{\Sigma_{\rm e}(\Delta+i\gamma)\} \simeq \frac{|g|^2 \gamma}{2\pi}\int^{+\pi}_{-\pi}dk \frac{2+e^{ikd}+e^{-ikd}}{(\Delta+2J\cos{k})^2 + \gamma^2}. 
\end{align}
This integral is the same that the one calculated in Eq.~\eqref{eqSM:integral 2 pumps I_T} to compute the total intensity of a 1D lattice driven by two pumps. Therefore, we get
\begin{align}
    \Gamma_{\rm e}(\Delta) = 4\pi|g|^2 D(\Delta)
    \left[ 1 + \cos(k_0 d) e^{-\pi\gamma D(\Delta)d} \right].
\label{eqSM:LDOS 1D}
\end{align}
The above expression follows the same functional dependence of Eq.~\eqref{eqSM:2 pumps I_T}, but making the substitution $g\leftrightarrow F/\sqrt{2\gamma}$. 

Therefore, the light-matter coupling in the quantum emitteres picture plays the role of the coherent pump amplitudes in the driven-dissipative setup. Although this was already demonstrated by Ref.~\cite{Gonzalez-Tudela_2022}, the connection between the total intensity and the LDOS that we have unveiled in this Section was, to our knowledge, never explored in the literature. Such a connection can be employed to measure the LDOS by means of the total intensity, which is more experimentally accessible. Besides, the correspondence between these two quantities allows us to regard the vanishing total intensity which accompanies a high localization parameter as an inefficient coupling of the coherent drivings to the photonic lattice.

\subsection{2D lattice}
\label{secSM:LDOS 2D}

\begin{figure}[tb]
    \centering
    \includegraphics[width=0.3\linewidth]{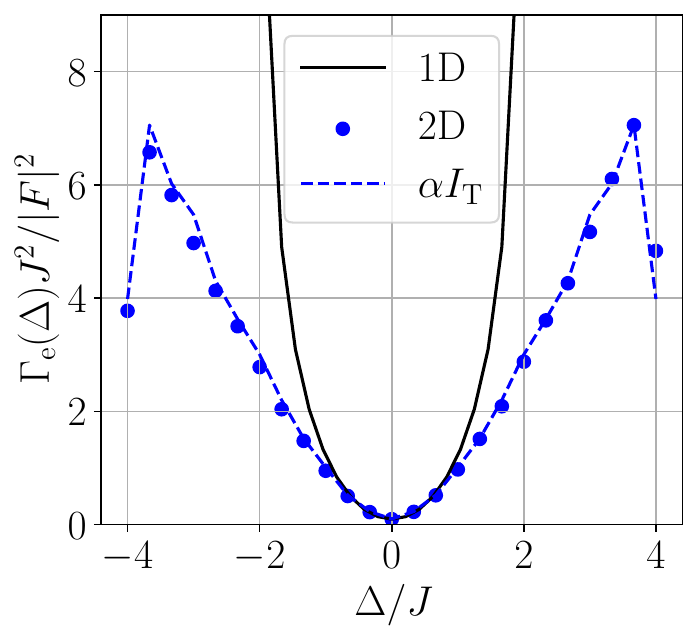}
    \caption{Black solid line: LDOS $\Gamma_{\rm e}$ (normalized in units of the tunneling rate $J$ and the pump intensity $|F|^2$) for two quantum emitters coupled at two different sites separated by a distance $d=2$ in a 1D lattice. Blue circles: LDOS for four quantum emitters coupled to sites $(\pm 1,0)$ and $(0,\pm 1)$ of a 2D square lattice. Both LDOS are plotted as a function of the detuning $\Delta$. Blue dashed line: total intensity $I_{\rm T}$ in the driven-dissipative 2D lattice [under coherent pumping at sites $(\pm 1,0)$ and $(0,\pm 1)$] multiplied by a proportionality factor $\alpha\simeq 0.11$. Simulation parameters: cavity decay rate $\gamma=5\times 10^{-2}J$, number of sites $100\times 100$ to calculate the LDOS of the square lattice and $50\times 50$ to compute its $I_{\rm T}$.}
    \label{figSM:LDOS 1D 2D}
\end{figure}

We now consider a 2D square lattice with $N\times N$ sites and first-neighbor couplings. Four quantum emitters are coupled sites $(\pm 1,0)$ and $(0, \pm 1)$, where the inter-site distance is $1$ and $(0,0)$ is the central site. The self-energy of the quantum emitters reads
\begin{align}
    \Sigma_{\rm e}(\Delta+i\gamma) = \frac{1}{N^2} \sum_{k_x,k_y} \frac{|g e^{-i k_{\rm x} d} + g e^{i k_{\rm x} d} + g e^{-i k_{\rm y} d} + g e^{i k_{\rm y} d}|^2}{\Delta + 2J\left[ \cos{k_{\rm x}} + \cos{k_{\rm y}} \right] + i\gamma }.
\label{eqSM:self energy 2D}
\end{align}
Although it is difficult to analytically perform the calculation above, even by transforming the sum into an integral, we can still calculate the the self energy numerically. As in the 1D case, the LDOS is then evaluated as $\Gamma_{\rm e} = -2\rm{Im}\{\Sigma_{\rm e}(\Delta+i\gamma)\}$.

The resulting LDOS is shown in  Fig.~\ref{figSM:LDOS 1D 2D}, where we compare it with the total intensity $I_{\rm T}$ of a driven-dissipative square lattice with four coherent pumps located at the same positions as the emitters, as was studied in Sec.~5 of the Main Text. The total intensity is rescaled using a proportionality factor $\alpha\simeq 0.11$ in order to show that, as in the 1D case, the LDOS of the square lattice is proportional to $I_{\rm T}$.

Finally, Fig.~\ref{figSM:LDOS 1D 2D} also compares the LDOS of the square lattice with the one of the 1D lattice [given by Eq.~\eqref{eqSM:LDOS 1D}]. As we can see, the main difference between them is that the LDOS of the 1D lattice diverges at its band edges ($\Delta=\pm 2J$), while that of the square lattice reaches a constant value at the corresponding band edges ($\Delta = \pm 4J$). We note that this provides some intuition behind the fact that the dynamical instability observed in the calculations shown in Figs.~3 and 7 of the Main Text appears for a smaller value of detuning in the 1D lattice compared to the 2D one. As a larger number of states (given by the diverging LDOS) is available in the former situation, parametric terms coupling different momenta are more important, and thus drive the dynamical instability sooner than when the 2D lattice is considered. More information on the role of parametric terms is provided on the following Section of this Supplement.

\section{Dynamical instability in momentum space}

\begin{figure}[tb]
    \centering
    \includegraphics[width=0.3\linewidth]{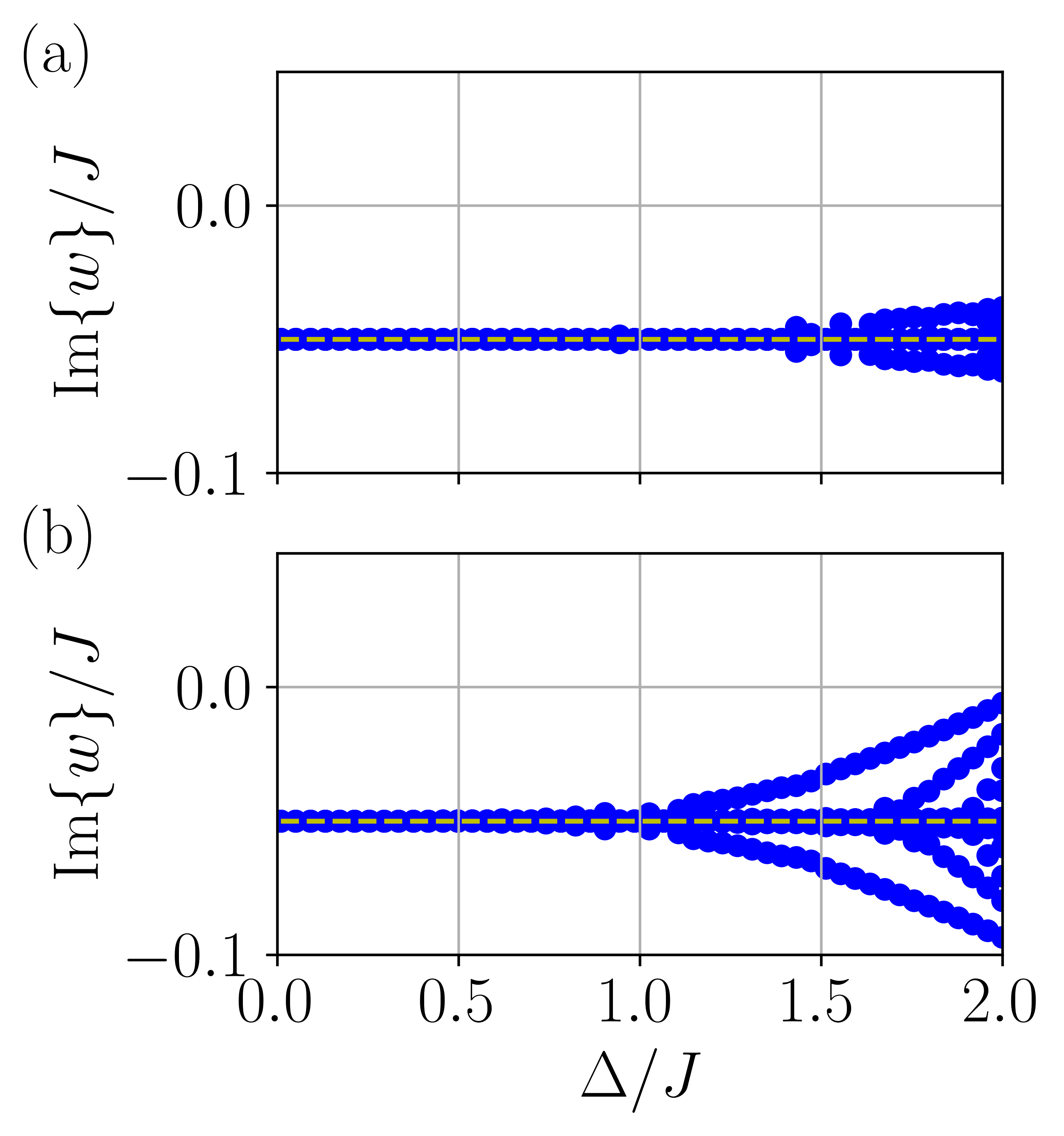}
    \caption{Dynamical stability in the 1D lattice coherently driven by two pumps separated by a distance $d=2$. All panels show the imaginary part of the eigenvalues Im$\{w\}$ of the Bogoliubov matrix $A$ (see the text) as a function of the detuning $\Delta$ (both quantities are normalized in units of the tunneling rate $J$) along the $\Delta=U|F|^2/J^2$ path, where $U$ is the Kerr non-linearity strength and $|F|^2$ is the pump intensity. (a) Self-Kerr interaction. (b) First-neighbors Kerr interaction. In all panels, the yellow dashed line represents the linear solution Im$\{w\}=-\gamma$. Simulation parameters: cavity decay rate $\gamma=5\times 10^{-2}$, $N=500$ sites with open boundary conditions, $n_1=250$.}
    \label{figSM:instability}
\end{figure}

In this Section we perform a Bogoliubov analysis of the dynamical stability in the 1D lattice. Our intuition is that the parametric terms coupling different momenta give rise to dynamical instability for large values of detuning $\Delta$ and pump intensity $|F|^2$, thus destroying localization. We consider a driving configuration with two pumps separated by a distance $d=2$.

We begin by transforming the non-linear coupled-mode equations~(1) of the Main Text into reciprocal space:
\begin{align}
    i\dot{a}_k = -(\Delta+2J\cos{k})a_k
    + \frac{U}{N}\sum_{q,q'}a^{*}_{q+q'-k}a_q a_{q'}
    -i\gamma a_k
    +F_k,
\label{eqSM:full coupled mode k}
\end{align}
where $a_k = (1/\sqrt{N})\sum_n a_n e^{-ikn}$ and $F_k = (1/\sqrt{N})\sum_n F_n e^{-ikn}$ are the Fourier transforms of the electric field amplitude and the coherent driving, and the lattice momentum $k$ runs from $-\pi$ to $\pi$ with steps $2\pi/N$. Parametric terms appear due to the Kerr non-linearity, and they couple different momenta while satisfying the conservation law for such quantity.

In principle, we need to take into account all terms appearing in the double sum of Eq.~\eqref{eqSM:full coupled mode k}. However, to study the appearance of instability in more detail we start considering independently all the different coupling terms in $k$. Although these approximations are by no means justified (a localized intensity in real space results in the population of all momenta in reciprocal space), we use them as a first step to understand better the role of the different parametric terms.

We start by considering the self-Kerr interaction, which results in the coupled-mode equations
\begin{align}
    i\dot{a}_k = -(\Delta+2J\cos{k})a_k
    + \frac{U}{N} |a_k|^2 a_k
    -i\gamma a_k
    +F_k.
\label{eqSM:motion self kerr}
\end{align}
In order to study the stability of small fluctuations around the steady state, we make the substitution $a_k\rightarrow a^{\rm (ss)}_k + \delta a_k$ in Eq.~\eqref{eqSM:motion self kerr}, where $a^{\rm (ss)}_k$ is the steady state reached at $t\rightarrow\infty$ and $\delta a_k$ are the small fluctuations around it. We then linearize the resulting equations with respect to $\delta a_k$, obtaining the Bogoliubov coupled-mode equations for the small fluctuations
\begin{align}
    i\delta\dot{a}_k &= -(\Delta+2J\cos{k})\delta a_k + 2U|a^{\rm (ss)}_k|^2\delta a_k + U {a^{\rm (ss)}_k}^2\delta a^*_k -i\gamma\delta a_k 
    \\
    i\delta\dot{a}^*_k &= (\Delta+2J\cos{k})\delta a_k - 2U|a^{\rm (ss)}_k|^2\delta a^*_k - U {a^{\rm (ss)}_k}^2\delta a_k -i\gamma\delta a_k 
    .
\end{align}
Using the vector form $\boldsymbol{\delta a}=[\delta a_{-\pi},\dots ,\delta a_{\pi}]^{\rm T}$ we can write the linear system of equations above in the more compact form
\begin{align}
    i\frac{d}{dt}\boldsymbol{\delta a}
    = A \boldsymbol{\delta a}.
\label{eq:Bogoliubov}
\end{align}
To assess the dynamical stability of the steady state against small fluctuations, one can diagonalize the matrix $A$ and look at the sign of the imaginary part of its eigenvalues Im$\{w\}$. Dynamical instability arises whenever there is at least one eigenvalue featuring a positive imaginary part~\cite{CarusottoCiuti2013,MunozDeLasHeras2021}.

Im$\{w\}$ is shown in Fig.~\ref{figSM:instability}(a) as a function of $\Delta$ across the path $\Delta=U|F|^2/J^2$ where localization appears. For comparison, we also show Im$\{w\}$ in the linear regime (i.e., by setting $U=0$) as a yellow dashed line. As expected, in the linear case  Im$\{w\}=-\gamma$ regardless of $\Delta$, meaning that the system can never become dynamically unstable. However, when a finite value of $U$ is added, the self-Kerr interaction makes the imaginary part of some eigenvalues depart from their linear values, effectively broadening Im$\{w\}$. This takes place near the upper band edge, located at $\Delta=2J$, which agrees with the fact that dynamical instabiltiy is observed for growing $\Delta$ when all terms in Eq.~\eqref{eqSM:full coupled mode k} are taken into account. However, as we see from Fig.~\ref{figSM:instability}(a), dynamical instability cannot be induced by self-Kerr terms alone.

The next order in our expansion are parametric terms coupling neighboring momenta (i.e., each value of $k$ with its corresponding $k\pm 2\pi/N$). The coupled-mode equations read in that case
\begin{align}
    i\dot{a}_k &= -(\Delta+2J\cos{k})a_k
    + \frac{U}{N}|a_k|^2 a_k
    + 2\frac{U}{N}|a_{k+2\pi/N}|^2 a_k
    + 2\frac{U}{N}|a_{k-2\pi/N}|^2 a_k \nonumber\\
    &+ 2\frac{U}{N} a^{*}_{k}a_{k+2\pi/N}a_{k-2\pi/N}
    -i\gamma a_k
    +F_k.
\label{eqSM:motion neighbor kerr}
\end{align}
An analog Bogoliubov analysis can be performed in this case as well. The imaginary part of the eigenvalues of the matrix $A$ governing the dynamics of linearized fluctuations is shown in Fig.~\ref{figSM:instability}(b) as a function of $\Delta$ across the $\Delta=U|F|^2/J^2$ path. Even though the system is still dynamically stable along the whole path, some imaginary parts depart from the linear solution Im$\{w\}=-\gamma$ sooner than when only self-Kerr terms are present, and they even reach Im$\{w\}=0$ at the upper band edge (i.e., at $\Delta=2J$). 

In conclusion, parametric couplings between different values of $k$ result in some Im$\{w\}$ departing from their linear value and approaching zero. Our intuition is that the inclusion of the rest of the parametric terms in Eq.~\eqref{eqSM:full coupled mode k} results in the appearance of a dynamical instability near the upper band edge, which is what we observe when we solve the coupled mode equations in real space~(1)  of the Main Text, as it shown in Figs.~3 and 7.

\end{widetext}

\end{document}